\documentstyle[epsfig]{mn}

\def\msun{\rm M_{\odot}}

\def\simlt{\mathrel{\rlap{\lower 3pt\hbox{$\sim$}}\raise 2.0pt\hbox{$<$}}}

\def\simgt{\mathrel{\rlap{\lower 3pt\hbox{$\sim$}} \raise 2.0pt\hbox{$>$}}}

\def\lsim{\mathrel{\rlap{\lower 3pt\hbox{$\sim$}}\raise 2.0pt\hbox{$<$}}}

\def\gsim{\mathrel{\rlap{\lower 3pt\hbox{$\sim$}} \raise 2.0pt\hbox{$>$}}}

\def\mbulge{M_{\rm Bulge}}

\def\msunpc3{\msun~{\rm {pc^{-3}}}}

\newcommand{\be}{\begin{equation}}

\newcommand{\ee}{\end{equation}}

\begin{document}

\title[Dual massive black holes]{Dual black holes in merger remnants. I: linking accretion to dynamics.}
\author[Dotti et al.]{M.~ Dotti$^{1,}$$^2$\thanks{e-mail address:
    mdotti@umich.edu}, M.~Ruszkowski$^{1,}$$^3$, L.~Paredi$^2$, M.~Colpi$^4$,
  M.~Volonteri$^1$, F.~Haardt$^2$\\
$^1$ Department of Astronomy, University of Michigan, Ann Arbor, MI, 48109, USA\\
$^2$ Dipartimento di Fisica e Matematica, Universit\`a  dell'Insubria, Via Valleggio 11, 22100 Como, Italy\\
$^3$ The Michigan Center for Theoretical Physics, Ann Arbor, MI, 48109, USA\\
$^4$ Dipartimento di Fisica G.~Occhialini, Universit\`a degli Studi di Milano
Bicocca, Piazza della Scienza 3, 20126 Milano, Italy\\
}
\maketitle \vspace {7cm}

\begin{abstract}

We study the orbital evolution and accretion history of massive
black hole (MBH) pairs in rotationally supported circumnuclear
discs up to the point where MBHs form binary systems. 
Our simulations have high
resolution in mass and space which, for the first time, makes it feasible to 
follow the orbital decay of a MBH either counter-- or co--rotating with
respect to the circumnuclear disc. We show that 
a moving MBH on an initially counter--rotating orbit experiences an
``orbital angular momentum flip'' due to
the gas-dynamical friction, i.e., it starts to corotate with
the disc before a MBH binary forms. We stress that this effect 
can only be captured in very high resolution simulations. Given the extremely
large number of gas particles used, the dynamical range is sufficiently large to 
resolve the Bondi-Hoyle-Lyttleton radii of individual MBHs.
As a consequence, we are able to link the accretion processes to the
orbital evolution of the MBH pairs.
We predict that the accretion rate is significantly 
suppressed and extremely variable when the MBH is moving on a retrograde orbit.
It is only after the orbital angular momentum flip has taken place that 
the secondary rapidly ``lights up'' at which point both
MBHs can accrete near the Eddington rate for a few Myr. The  
separation of the double nucleus is expected to be around $\lsim 10$ pc at this stage.
We show that the accretion rate can be highly variable also 
when the MBH is co--rotating with the disc (albeit to a lesser extent) provided that 
its orbit is eccentric.
Our results have significant consequences for the expected number of
observable double AGNs at separations of $\lsim 100$ pc.
\end{abstract}


\section{Introduction}

 In recent years, compelling observational evidence has indicated
that
 massive black holes (MBHs) are ubiquitous in nuclei of local bright
galaxies (Richstone et al. 1998; Marconi et al. 2004; Shankar et al. 2004;
Ferrarese et al. 2006; Ferrarese 2006; Decarli et al. 2007).  According to
the structure formation paradigm, galaxies often interact
 and merge as their
dark matter halos assemble in a hierarchical
 fashion, and the MBHs,
incorporated through mergers, are expected to
 grow, evolve and {\it pair}
with other MBHs (Begelman Blandford \&
 Rees 1980; Volonteri, Haardt, \&
Madau 2003).  The formation of
 MBH pairs thus appears to be an inevitable
and natural
 consequence of galaxy assembly.
 
 A large number of merging
systems such as the luminous infrared
 galaxies (LIRGs) hosts a central
rotationally supported massive (up to
 $10^{10}\msun$) gaseous disc extending
on scales of $\sim 100$ pc
 (Sanders \& Mirabel 1996; Downes \& Solomon 1998;
Davies et al.
 2004). These discs may be the end--product of
gas--dynamical,
 gravitational torques excited during the merger, when a
large amount of
 gas is driven into the core of the remnant (Mayer et
al. 2007).
 Inside a massive self--gravitating disc, a putative MBH pair can
continue
 its dynamical evolution and, possibly, accrete gas producing a
double AGN (Kocsis et al. 2005, Dotti et al. 2006a).
 
 The possibility of
double accretion processes during the different
 stages of a galaxy merger is
still a matter of debate.  From an
 observational point of view, only few
tens of double quasars are known
 (with separations $\lsim 100$ kpc; see
e.g. Foreman, Volonteri \&
 Dotti 2008 and references therein).  There are
only a few well studied
 cases of ongoing mergers at sub-galactic scales in
which each
 companion is an AGN. Examples of such systems include NGC 6240
(Komossa et al. 2003), Arp 229
 (Ballo et al. 2004), and Mrk 463 (Bianchi et
al. 2008).
 Other two dual AGN candidates with separation $\sim 1$ kpc have
been identified
 spectroscopically by Comerford et al. (2008).  
 On parsec
scale, only one resolved active MBH binary has been found in the
 nucleus of
the elliptical galaxy 0402+369 (Rodriguez et al. 2006), and
 the existence of
two more sub--parsec MBH binary candidates has been 
 suggested (OJ287,
Valtonen et
 al. 2007, and SDSSJ092712.65+294344.0, Bogdanovic, Eracleous
\&
 Sigurdsson 2008; Dotti et al. 2008).

Accretion events during the last phases of a MBH binary (for
sub--parsec separations) have been studied in detail by several
authors (Armitage \& Natarajan 2002, 2005; Hayasaki, Mineshige, \&
Sudou 2007; Hayasaki, Mineshige, \& Ho 2008; Cuadra et al. 2009). On
larger scales, the possibility of gas accretion during the MBH pairing
has been recently studied using numerical simulations in the
context of galaxy mergers (e.g., Kazantzidis et al. 2005; Springel et
al. 2005; Di Matteo et al. 2005; Hopkins et al. 2005; Hopkins et
al. 2006).  In these works, the authors study a galaxy-galaxy
collision on spatial scales of $\sim 100$ kpc, and the number of
particles used to model the two galaxies does not allow them to
resolve the influence radii of the two pairing MBHs. This lack of
resolution prevents the authors from accurately predicting the
accretion rates onto the two MBHs. Given the limits of the currently
available supercomputers, such simulations would be prohibitively
time
 consuming. Therefore, no simulations resolving the
Bondi-Hoyle-Lyttleton radii during a complete
 galaxy--galaxy merger
have been published to date.
 
 We have run a suite of high
resolution N--body/hydrodynamical
 simulations of MBH pairing in
circumnuclear discs. 
 In this paper we discuss the
 dynamical
evolution of the MBH pairs up to a point where they form a binary
(i.e., up to a point where the mass enclosed inside the MBH orbit is
smaller than the sum of the 
 MBH masses). We follow the MBH
dynamics from an initial separation of 50 pc down to sub--parsec
separation while resolving $R_{\rm BHL}$ of the two pairing
 MBHs. We
predict the accretion rate onto the two MBHs as a function of 
 the
geometry of the merger.
 As we demonstrate in this paper, this
approach allows us to study 
 the possibility that a galaxy merger
event may be associated with 
 a single or double active galactic
nucleus
 depending on the configuration of the 
 initial MBH
orbits. It also allows us 
 to constrain the fate of the MBH binary
(i.e., coalescence
 vs. stalling).  
 The evolution of the MBH spins
and the fate of the newly formed MBH binary
 will be discussed in two
forthcoming papers.

\section{Simulation setup}

We follow the dynamics of MBH pairs in nuclear discs using numerical
simulations run with the N--Body/SPH code GADGET (Springel, Yoshida \&
White 2001), upgraded to include the accretion physics.  The main
input parameters of our simulations are summarized in Table~1.

\begin{table}
\label{tab:run}
\begin{center}
\caption{Run parameters}
\begin{tabular}{l@{   }c@{   }c@{    }c@{    }c@{   }}  
\hline
\\
run & ~~~prograde ? & ~~~$e_{i}$ & ~~~$\gamma$  \\
\\
\hline
\hline 
\\
HPC & ~~~yes &  ~~~0    &     \\
HPE & ~~~yes &  ~~~0.7  & ~~~5/3, ``hot'' \\
HRE & ~~~no  &  ~~~0.7  &     \\ 
\hline
CPC & ~~~yes & ~~~0     &     \\
CPE & ~~~yes & ~~~0.7   & ~~~7/5, ``cold'' \\
CRE & ~~~no  & ~~~0.7   &     \\
\hline
\end{tabular}\\
\end{center}
\noindent
\end{table}

The initial conditions of our runs are set following the procedure
discussed in Dotti, Colpi \& Haardt (2006b) and Dotti et al. (2007).
In our models, two MBHs are placed in the plane of a gaseous disc,
embedded in a larger stellar spheroid.  The gaseous disc is
modeled with 2,353,310 particles, has a total mass $M_{\rm{Disc}}=10^8
\msun$, and follows a Mestel surface density profile 
\be
\Sigma(R)=\frac{M_{\rm disc}}{2 \pi R_{\rm disc} R} 
\ee
 where $R$ is the radial distance projected into the disc plane and
$R_{\rm disc}= 100$ pc is the  size of the disc.  The disc
is rotationally supported in $R$ and has a vertical thickness
of 10 pc. Initially, the gaseous particles are distributed uniformly
along the vertical axis.  The internal energy per unit mass of the
SPH particles scales as: 
\be 
u(R)=K R^{-2/3}, 
\ee 
where $K$ is a constant defined so that the Toomre parameter of the
disc,
\be
\label{Toomre} Q=\frac{\kappa c_{\rm s}}{\pi G \Sigma}, 
\ee
is $\geq 3$ everywhere, preventing the fragmentation of the
disc (the average value of $Q$ over the disc surface is $\approx
10$).  In
 equation~(\ref{Toomre}) $\kappa$ is the local epicyclic
frequency, and
 $c_{\rm s}$ is the local sound speed of the gas.  Gas
is evolved
 adiabatically assuming a polytropic index $\gamma=5/3$ or
$\gamma=7/5$. 
 In the former case, the runs are denoted by ``1'' and
are 
 termed ``hot'' as the temperature is proportional to a higher
power of density
 than in the latter class of runs (``cold'' cases,
runs denoted by ``2''). 
 The hot case
 corresponds to adiabatic
monoatomic gas, while the cold has 
 been shown to provide a good 
approximation to a solar metallicity gas heated by a starburst
(Spaans \& Silk 2000; Klessen, Spaans, \& Jappsen 2007).
 
 The
spheroidal component (bulge) is modeled with $10^5$ collisionless
particles, initially distributed as a Plummer sphere with mass
density
 profile
 \be 
 \rho (r)={3 \over 4 \pi}{\mbulge\over b^3}
\left(1+{r^2\over b^2}\right)^{-5/2}, 
 \ee 
 where $b$ $(=50$ pc
$)$ is the core radius, $r$ the radial coordinate,
 and
$\mbulge(=6.98 M_{\rm{Disc}})$ is the total mass of the spheroid.
With such a choice, the mass of the bulge within $100$ pc is five
times the
 mass of the disc, as suggested by Downes \& Solomon
(1998).
 
 The two MBHs ($M_1$ and $M_2$) are equal in mass ($M_{\rm
BH}=4\times
 10^6\,\msun$). The initial separation of the MBHs is 50
pc.  $M_1$, called
 primary for reference, is placed at rest at the
centre of the
 circumnuclear disc. The secondary (denoted as $M_2$) 
is either corotating with the
gaseous disc on a circular orbit, or moving on an eccentric counterrotating or corotating 
orbit with respect to the circumnuclear disc.
 For
eccentric orbits, the initial ratio between the absolute values of the
radial and tangential velocity of $M_2$ is $v_{\rm rad}/v_{\rm
tan}=3$. This ratio sets the eccentricity of the first orbit of $M_2$
to $e\simeq
 0.7$.  Given the large masses of the disc and the
bulge, the dynamics of the
 moving MBH ($M_2$) is uneffected by the
presence of $M_1$ until the
 MBHs form a gravitationally bound
system. The subsequent evolution of the MBH binary will be
 discussed
in a forthcoming paper.
 
 We evolve our initial composite model
(bulge, disc and $M_1$)
 for $\approx 3$ Myrs, until the bulge and
the disc reach equilibrium,
 as described in Dotti et al. (2006b; 2007).  The disc, having initially finite thickness
and homogeneous vertical density distribution, is allowed to relax
to an equilibrium configuration along the z--axis. The new
distribution of the gas in the disc in the z direction becomes
non-uniform once the equilibrium configuration is reached. The new disc has a vertical
thickness of $\approx 8$ pc 
(defined as containing 90\% of the mass in the vertical direction at every radius)
This thickness is independent of the
radius and is $\approx 4/5$ of the initial un--relaxed value.

 We allow the gas particles to be accreted onto the MBHs if the
 following two criteria are fulfilled:\\ 
 
 $\bullet$ the sum of
the
 kinetic and internal energy of the gas particle is lower than
$\alpha$ times the absolute value of its gravitational energy (all the
energies are computed
 with respect to each MBH)\\ 
 \indent
$\bullet$ the total mass accreted per unit time onto a MBH every
timestep
 is lower than the accretion rate corresponding to the
Eddington luminosity ($L_{\rm Edd}$)
 computed assuming a radiative
efficiency ($\epsilon$) of 10\%.\\
 
 \indent
 The parameter
$\alpha$ is a constant defining how much a particle has
 to be bound
to a MBH to be accreted. We set $\alpha=0.3$. Note that
 due to the
nature of the above criteria, the gas particles can accrete
 onto the
MBHs only if the Bondi-Hoyle-Lyttleton radius, defined as 
 \be
\label{eq:BHL}
 R_{\rm BHL} = \frac{G\,M}{v_{\rm rel}^2+c_{\rm
s}^2},
 \ee
 is resolved in the simulations. In Eq.~(\ref{eq:BHL}),
$M$ is
 the mass of the MBH, $v_{\rm rel}$ is the relative velocity
between
 the gas and the moving MBH, and $c_{\rm s}$ is the sound
speed of the
 gas. $M$ is evolved by adding at each timestep
$1-\epsilon$ times the total mass of the accreted gas particles.
If the amount of bound gas particles implies super--Eddington
accretion, the code randomly selects a subsample of the particles
within $R_{\rm BHL}$ to be accreted, so as to prevent accretion from
reaching super-Eddington rates.  This simplified treatment of the
black hole feedback does not capture effects such as the 
accretion driven wind, and
its effect on the
environment. Because the MBHs accrete only a
small fraction of their initial masses ($< 10 \%$) during our simulations, 
AGN feedback is
not expected to modify the global properties of the circumnuclear
disc. Nevertheless, such feedback can remove a fraction of the gas
a few parsecs away from MBHs accreting near the Eddington limit. 
This could decrease the
accretion rate onto the MBHs and the efficiency of MBH pairing after
they form a binary system. The accurate implementation of a self--consistent 
radiative feedback is beyond the scope of this work, and
will be the subject of our future studies.

 The number of neighboring SPH particles adopted in our simulation is $N_{\rm
neigh}=32$.  The resulting spatial resolution of
the hydrodynamical force in the highest
 density regions is $\approx
0.1$ pc. In order to prevent numerical errors
 due to the
density-dependent effective resolution, we set both the
gravitational
 softening of the gaseous particles and that of the
MBHs to 0.1 pc.
 With this spatial resolution we can resolve the
vertical scale of the disc and, as a consequence, we can calculate angular momentum
transport due to disc self--gravity (see, e.g., Lodato \& Rice 2004;
Nelson 2006). We can also resolve the influence radius of $M_1$
($\approx 1$ pc) and $M_2$, a condition necessary to study the gas
accretion onto MBHs
\footnote{Note that $R_{\rm BHL}$ of $M_2$ depends on the phase of its
orbit. That is,
 $R_{\rm BHL}$ depends on the relative MBH--gas
velocity, which is different in 
 the corotating and counterrotating
cases.}. 
  
In addition, our accretion scheme is such that linear momentum is
conserved.  This includes automatically the braking exerted by the
accretion flow onto the moving MBH (${\bf F}_{\rm BHL}=\dot{M} 
\,{\bf v}_{\rm rel}$, where $\dot{M}$ is the accretion rate).
The
only viscosity term in the simulations is the shear reduced version (Balsara 1995;
Steinmetz 1996) of the standard Monaghan and Gingold (1983) artificial
viscosity.

\section{RESULTS}

\subsection{Orbital evolution}

In this Section we discuss the orbital evolution of the pairs in
six different runs, until the MBHs form a binary. The relationship
between the orbital properties of the MBHs and their accretion
properties is discussed in Section~\ref{sec:acc}. The run names and parameters 
are listed in Table 1. The letters used to name the runs encode the type of the run
with the first, second and third letter corresponding to 
hot vs. cold, prograde vs. retrograde and eccentric vs. circular cases, respectively. For example,
CRE corresponds to a secondary MBH on an eccentric retrograde orbit in a cold disk.

\subsubsection{Prograde orbits}

The pairing process of MBHs on circular orbits (runs HPC and CPC) brings
the two MBHs down to separations $\lsim 5$ pc in less than 10 Myrs at which point 
they form a binary. The MBHs orbital decay proceeds without any
observable eccentricity growth. Because of the low relative velocity
between the MBH and the gas, $M_2$ interacts efficiently with the gas out
to a distance of a few parsecs.  In this case, the dynamical evolution of the
pair is almost indistinguishable from lower (parsec scale) resolution
simulations (Escala et al. 2005; Dotti et al. 2006b; Dotti et al.
2007), before the MBHs bind in a binary.

\begin{figure}
\begin{center}
\centerline{\psfig{figure=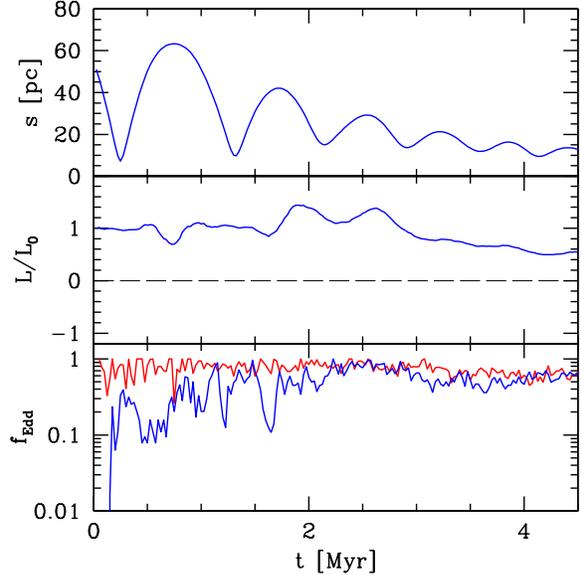,height=8cm}}
\caption{Run HPE. Upper panel: MBH separation as a function of time.
Middle panel: time evolution of the orbital angular momentum of $M_2$
($L$) normalized to its initial value ($L_0$). The thin--dashed
horizontal line marks $L = 0$.  Lower panel: Eddington accretion ratio
as a function of time. Red and blue lines refer to $M_1$ and $M_2$,
respectively.  }
\label{fig:APH}
\end{center}
\end{figure}

The upper panel of Fig.~\ref{fig:APH} shows the separation between the
two MBHs as a function of time for $M_2$ moving initially on a
prograde eccentric orbit inside the hot disc ($\gamma=5/3$, run
HPE). The two MBHs reach a separation of the order of $\lsim 10$ pc in
less than 5 Myr.  The MBH pair loses memory of its initial
eccentricity in the early phases of the orbital decay. Such
circularization is due to the variation in direction and in 
magnitude of dynamical friction acting on $M_2$ during an
orbital period.  Dynamical friction tends to slow down the
secondary near the periastron and to accelerate it at the apoastron
(see Dotti et al. 2006b; Dotti et al. 2007 for a detailed
discussion). The orbital decay and circularization observed in these
high resolution runs are slightly faster than in the low resolution
simulations presented in Dotti et al. (2006b) and Dotti et al.
(2007).  The increased resolution allows to better resolve the
interaction between the gas environment and $M_2$ when it is still
moving on eccentric orbits and the relative velocity between $M_2$ and
the gas particles is high. This effect of the increased resolution is
more prominent for retrograde orbits, where the relative velocities
are higher. We postpone a detailed discussion of this effect to
Section 3.1.2.

The thermodynamics of the gas also has impact on the circularization
process.  The dynamical interaction of the moving MBH of mass $M_2$
with the gas is stronger in a cold disc.  This is especially so when
the secondary is corotating with the disc. The reason for this is that
the relative velocity between the secondary and the disc material is
small in this case (at least for circular orbits) and the
Bondi-Hoyle-Lyttleton radius reduces to the Bondi radius which is
proportional to $c_{s}^{-3}$, where $c_{s}$ is the sound speed. It is
for this reason that the orbital decay and the circularization in a
cold disc (run CPE) proceeds faster than in the hot case.
In our simulations dynamical friction is
efficient in reducing the eccentricity down to values $<0.1$, for both
cold and hot cases.

\subsubsection{Retrograde orbits}

\begin{figure}
\begin{center}
\centerline{\psfig{figure=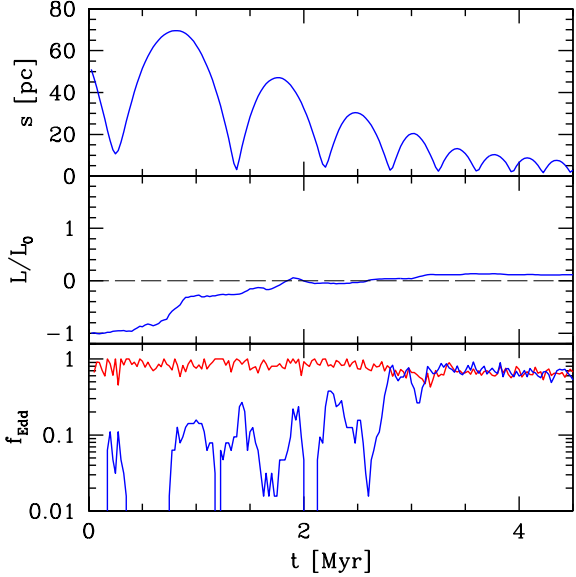,height=8cm}}
\caption{Same as Fig.~\ref{fig:APH} for run HRE. In the middle panel the initial
angular momentum is considered negative since $M_2$ is initially
orbiting on a retrograde orbit.
}
\label{fig:ARH}
\end{center}
\end{figure}

\begin{figure}
\begin{center}
\centerline{\psfig{figure=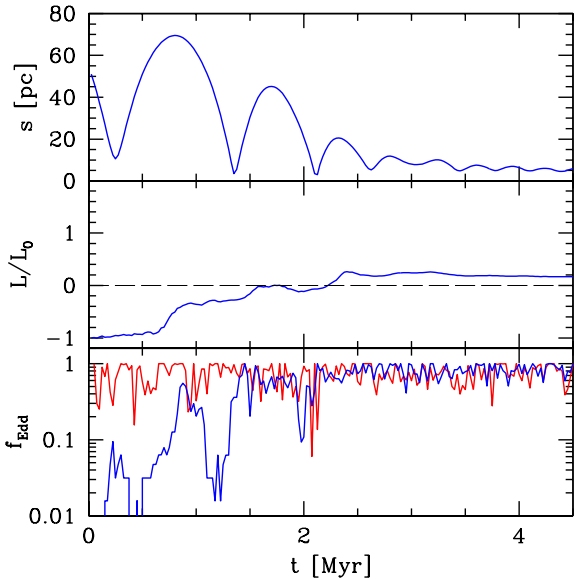,height=8cm}}
\caption{Same as Fig.~\ref{fig:APH} and Fig.~\ref{fig:ARH} for run CRE.}
\label{fig:ARC}
\end{center}
\end{figure}

 For the retrograde cases (runs HRE and CRE) we find that the two
MBHs also reach
 a separation of $\lsim 10$ pc in less than 5 Myr, as
illustrated in
 the upper panels of Fig.~\ref{fig:ARH} and
Fig.~\ref{fig:ARC}.  The
 middle panels of Fig.~\ref{fig:ARH} and
Fig.~\ref{fig:ARC} show the
 evolution of the z--component of the
orbital angular momentum $L_{\rm
 z}$ of $M_2$ normalized to its
initial value. The (initially negative) angular momentum
 grows very
fast during the first Myr. As long as $M_2$ is counter--rotating,
the
 MBH--disc interaction brakes the secondary even at the
apocentre,
 because the MBH velocity and disc flow are anti--aligned
there.
 The increase of the orbital angular momentum of $M_2$ is
further facilitated by 
 the fact that, while the orbit decays, the
MBH interacts with progressively denser 
 regions of the disc closer
to the primary. Eventually, at about 2 Myr, the sign of the orbital 
momentum of the secondary changes.
 The dynamical friction process is
the ultimate cause of this orbital
 angular momentum flip.  
 After
the orbital angular momentum flip has taken place 
 and the secondary
enters a co--rotating orbit with respect to the disc,
 the angular
momentum continues to grow up to $L/L_0\,\approx 0.1$. 
 
 The
orbital decay of a retrograde MBH (runs HRE and CRE), and the
orbital angular momentum flip process are dependent on
 the numerical
resolution.  We stress that 
 the orbital angular momentum flip is
not seen in lower resolution
 simulations. This resolution dependence

 has a simple dynamical explanation. 
 The perturbation of the
orbit of a single gas particle
 that is caused by $M_2$ depends on
the relative
 velocity between the gas and the MBH and their relative
separation.
 Only those gas particles that get perturbed can
contribute to 
 the orbital decay of the MBH and lead to its orbital
angular momentum flip.
 A natural scale lenght for the gravitational
interaction 
 between the MBH and gas particles is $R_{\rm
 BHL}$.

 Gas particles can efficiently exchange angular
 momentum with
the hole only if they pass closer to the MBH than a few $R_{\rm BHL}$.

 As long as the MBH is counter--rotating, its relative
 velocity
with respect to the disc is $v_{\rm rel}\gsim 200$ km s$^{-1} \gg
c_s$, which corresponds to $R_{\rm BHL}\lsim 1$ pc.  
 Therefore,
higher numerical resolution is required to model counter--rotating
cases
 than the co--rotating ones. Given the high
 resolution of our
simulations, the gravitational interaction is
 accurately computed
down to scales of $\approx 0.1$ pc, allowing us to
 correctly model
the dynamical evolution of counter--rotating MBHs.

\subsection{Accretion history}\label{sec:acc}

 In run HPC the Eddington ratio $f_{\rm Edd}\equiv \dot{M} /
\dot{M}_{\rm Edd}$ of the primary is $\approx 1$ at the beginning of
the simulation and decreases monotonically during the pairing
process
 down to $\approx 0.7$ at 5 Myr. The Eddington ratio of the
primary
 decreases steadily because the orbital motion of the
secondary heats
 up the gas in the vicinity of the primary. The
increase in the gas
 temperature decreases the Bondi radius of the
primary and evacuates 
 gas from the central regions of the
disc. Both of these effects reduce
 the accretion rate onto the
primary. This effect is present only when
 the polytropic index
$\gamma= 5/3$, as in that case the increase of the
 gas temperature
with density is larger than in the cold case where
 $\gamma= 7/5$
(run CPC).  In the cold run, the lower value of
 $\gamma$ allows for
the pairing of the MBHs while preventing a strong
 heating of the
gas.  
In
 both runs (HPC and CPC), $M_2$ has an
accretion history similar to
 $M_1$. That is, MBHs on circular
corotating orbits have low relative
 velocities with respect to the
gas which leads to high and steady
 accretion rates.
 
 Figures
~\ref{fig:APH}, \ref{fig:ARH} and \ref{fig:ARC} allow to
 compare
directly the dynamical properties of the two MBHs to the
 accretion
rates in runs HPE, HRE, and CRE, respectively. Top panels show
 the
evolution of the MBH separation. Middle panels present the time
dependence of the orbital angular momentum of $M_2$.  The bottom
panels show $f_{\rm Edd}$ for the primary (red lines) and secondary
(blue lines) MBHs.
 
 In runs HPE and HRE, the primary accretes at
an average $f_{\rm Edd} \approx
 0.7$, slightly decreasing with
time. The accretion rate of the secondary
 evolves differently. Its
accretion behaviour depends on
 whether it is on a co--rotating or
counter--rotating orbit.
 In the co--rotating case (run HPE), the
average $f_{\rm Edd}$ of the
 secondary is $\approx 0.4$ during the
first 5 Myr and its accretion
 history can be divided into two
phases:\\
 \indent
 i) for $t\lsim 2.5$ Myr the
 circularization
process is efficient while $f_{\rm
 Edd} \approx 0.3$ on average and
shows strong variability\\ 
 \indent
 ii) for $t\gsim 2.5$ Myr,
$M_2$ moves on a quasi--circular orbit, and the
 relative velocity
between $M_2$ and the gaseous disc is reduced. In
 this phase,
$f_{\rm Edd} \approx 0.45$.\\ 
 
 In the counter--rotating case (run
HRE), we can still distinguish two phases 
 albeit with some
important differences:\\
 \indent
 i) for $t\, \lsim \, 3$ Myr,
$M_2$ is counter--rotating
 ($L_{\rm z} < 0$), and $f_{\rm Edd}\sim
0.05\div 0.1$. 
 This level of accretion is lower than in the
corresponding 
 first orbital stage in the co--rotating case
described above. Moreover, the variability in the accretion rate 
onto the secondary MBH is now significantly larger.\\ 
 \indent
 ii)
At $t\sim 3$ Myr the orbital angular momentum flip takes place.
Thereafter, $M_2$ accretes at $f_{\rm Edd} \sim 0.7$. 
 In this
second stage, the main difference between the co--rotating and
 the
counter--rotating cases is that, in the latter, $M_2$ exhibits a
rapid transition/increase in the accretion rate (cf. bottom panels
in
 Fig. 2 (counter--rotating case) and Fig 1 (co-rotating
case). In the next subsection we present the Fourier analysis of
the accretion history of $M_2$, in order to
assess whether the variability of $f_{\rm Edd}$ is physical or it depends
on numerical fluctuations due to finite number of particles used.

In run CRE, $M_2$ has a similar accretion history to the one
observed
 in run HRE, but the timescale for the orbital angular
momentum flip is
 shorter. Lower $\gamma$ corresponds to colder and
denser gas and,
 thus, more efficient dynamical friction. This makes
$M_2$ corotate in
 less than 2 Myr. The primary has an average
$f_{\rm Edd} \approx 0.9$,
 which is higher than in runs HPE and HRE.
The physical reason for the
 enhancement in the accretion rate in the
cold case is provided in
 the discussion.

\subsubsection{Fourier analysis of accretion fluctuations in run HRE}

As discussed above, strong fluctuations of $f_{\rm Edd}$ are a
distinctive feature of counter--rotating (and eccentric co--rotating)
MBHs.  The variability is higher for counter--rotating MBHs and
$f_{\rm Edd}<0.1$ (see Fig.~\ref{fig:ARH} and \ref{fig:ARC}). These
values of $f_{\rm Edd}$ correspond to less than a few hundred SPH
particles accreted every $2.5 \times 10^4$ yr which is the time resolution in
Fig.~\ref{fig:ARH} and \ref{fig:ARC}. Such a small number of
particles is $\lsim 10 \times N_{\rm neigh}$, and can drop down
to 0 when the $R_{\rm BHL}$ of $M_2$ is not resolved. As a
consequence, values of $f_{\rm Edd}<0.1$ should be considered order of
magnitude estimates.

In order to check whether the variability of $f_{\rm Edd}$ is due to
finite sampling of the medium or if it has a physical meaning, we
performed the Fourier analysis for run HRE\footnote{In run CRE $M_2$
is counter--rotating for less than 2 orbits, so the Fourier analysis
in this case would not be meaningful.}. We emphasise that run HRE is
the run with the lowest $f_{\rm Edd}$, i.e., most affected by the
numerical errors. Therefore our conclusions will be conservative and
robust.

\begin{figure}
\begin{center}
\centerline{\psfig{figure=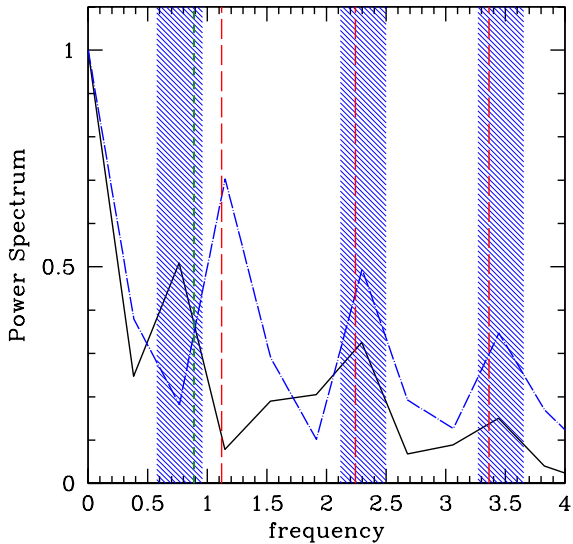,height=8cm}}
\caption{Power spectrum $P$ of $f_{\rm Edd}$ fluctuations of $M_2$ in
 run HRE, until the MBH is counter--rotating with respect to the
 circumnuclear disc. Solid black line: $P$ of the original $f_{\rm
 Edd}$. Dot--dashed blue line: $P$ of the larger scale/noise free
 estimate of the $f_{\rm Edd}$ discussed in the text.  $P$ has been
 normalized to its value at $\nu=0$, $\nu$ is in units of the average
 between $\nu_1=1/\tau_1$ and $\nu_2=1/\tau_2$. Short dashed green line
 marks $\nu=\nu_1$, long dashed red lines shows $\nu=\nu_2$ and its
 higher harmonics. The dashed regions show the values of $\nu$
 compatible with the peaks of $P$ (i.e. the frequencies of the peaks
 $\pm \,1/2\, \nu_{\rm min}$).  }
\label{fig:pwr}
\end{center}
\end{figure}

The black solid line in Fig.~\ref{fig:pwr} refers to the power
spectrum ($P$) of the fluctuations of $f_{\rm Edd}$ of $M_2$, for run
HRE.  The frequency ($\nu$) unit is defined as the average between
$\nu_1=1/\tau_1$ and $\nu_2=1/\tau_2$, where $\tau_1$ and $\tau_2$ are
the periods of the first and second orbit of $M_2$. $P$ has been
normalized to $P(\nu=0)$. The frequency resolution of our results is
$\nu_{\rm min}=1/\Delta t$, where $\Delta t \approx 3$ Myr corresponds
to the amount of time during which $M_2$ orbit is retrograde.
\footnote{The Nyquist critical frequency in the units of the
plot is $\approx 20$, well outside the range of $\nu$ discussed in
this section.}  

Fig.~\ref{fig:pwr} proves that $P$ is not compatible with white noise
as expected for fluctuations due only to numerical sampling. Moreover,
we can distinguish four peaks in our distribution.  The first peak at
$\nu = 0$ is a simple consequence of the physical condition $f_{\rm
Edd} \geq 0$. The remaining three peaks contain more physically useful
information. Given our frequency resolution, the frequency of the
second peak is compatible with $\nu_1$, while the third and fourth
peaks are at the frequencies corresponding to the second and third
harmonics of $\nu_2$ (i.e., $2\, \nu_2 $ and $3\,\nu_2$,
respectively). The natural interpretation of this result based on the
Fourier theorem is that $f_{\rm Edd}$ is the composition of two
periodic
signals, with periodicities $\nu_1$ and $\nu_2$. 

As an additional proof for the ``physical'' nature of the variability
of $f_{\rm Edd}$, we computed an ``a posteriori'' estimate of $f_{\rm
Edd}$ by averaging the gas properties in the vicinity of the secondary
every $2.5 \times 10^4$ yr and then the using these averages to
estimate the accretion rates. This approach allowed us to eliminate
the effects of the sampling noise due to a finite number of SPH
particles.  More specifically, we computed the mean values of the
density, relative velocity, and temperature of the gas inside the
sphere of radius $R_{\rm p}=7$ pc centered around $M_2$.  The choice
of the radius $R_{\rm p}$ is motivated by two arguments: $R_{\rm p}$
is small enough to resolve the substructures in the disc caused by the
orbiting MBH, and large enough to have a statistically large number of
particles within $R_{\rm p}$ from $M_2$.  The average number of
particles inside the sphere defined by $R_{\rm p}$ is $\approx
20,000$, and the minimum number during the whole run is $\approx
2,000$ (i.e., $\approx 60 \times N_{\rm neigh}$). Such large numbers
assure that the finite sampling of our disc does not affect our a
posteriori estimate.  Note that for counter--rotating $M_2$, $R_{\rm
p}> R_{\rm BHL}$. From these properties we estimated the
Bondi-Hoyle-Littleton accretion rate onto $M_2$.  The power spectrum
of the fluctuations of the new estimated $f_{\rm Edd}$ is shown in
Fig.~\ref{fig:pwr} as a blue dot--dashed line. The positions of the
second, third and fourth peaks are compatible with the $\nu_2$ and its
higher harmonics. The position of the first peak, different from the
one found for our original data, can be explained considering that
during the first phase of the first orbit $M_2$ has not yet developed
a parsec--scale overdensity in the gas. Consequently, $f_{\rm Edd}$
computed at large separations (i.e., within $R_{\rm p}$) during the
first orbit leads to lower signal in the frequency domain at $\nu_1$
compared to the one obtained considering only bound particles (black
curve).  Note that despite this (expected) difference, the spectrum
computed from the unaveraged data does possess a peak at $\nu=\nu_1$,
implying a physical origin of variability. 
\\ \indent In summary, the two tests discussed in
this section prove the ``physical'' nature of the $f_{\rm Edd}$
variability. The cause of such variability is discussed in the next
section.

\section{Discussion}

 For the first time, we are able to study the dynamical evolution
of
 MBHs on retrograde orbits embeded in circumnuclear disks.
These
 simulations require extremely high spatial resolution ($\lsim
0.1$
 pc). We discovered that dynamical friction acting on a MBH
moving on
 an initially retrograde orbit with respect to the disc
material leads
 to an orbital angular momentum flip. The moving MBH
always ends up on
 a prograde orbit by the time the MBH pair forms a
binary system
 (i.e., by the time the gas mass enclosed inside the
MBH orbit
 is smaller than the sum of the MBH masses).  We also find
that
 the circularization of the initially eccentric prograde orbit
of the
 secondary is efficient well before the formation of the
binary. We
 stress that the orbital angular momentum flip can only be
captured
 provided that the numerical resolution is sufficiently
high.
 
 The dynamical range of our simulations also allowed us to resolve the
 sphere of influence of the MBHs and thus, for the first time, study
 the mass accretion rate of the MBH co- and counter-rotating with
 respect to the disc.  We found that a highly variable double nuclear
 activity can be observed for few Myr when the two MBHs orbit each
 other with relative separations $\gsim 10$ pc. The accretion
 history of the moving MBH ($M_2$) can be divided into two distinct
 phases. The first stage occurs when the MBH is either on an eccentric
 co--rotating orbit or on a retrograde orbit irrespective of its
 eccentricity.  The second stage occurs after the circularization of
 the orbit or after the orbital angular momentum flip of $M_2$.  In
 the first stage, the accretion rate of the secondary on a
 counter--rotating orbit is more variable, and its luminosity lower by
 a factor of a few, than in a co--rotating case. In the second stage,
 the accretion rate onto the moving MBH is much enchanced.
 Interestingly, the accretion rate of a MBH that is initially on a
 counter--rotating orbit exhibits a significantly more rapid
 transition to the high-mass accretion rate, low-variability
 state. The luminosity increases by an order of magnitude.

We emphasise that the estimated accretion rate is computed on
scales resolved by the simulations. We assume that
the accretion rate is stationary.  The validity of this assumption
depends on the accretion disc physics on unresolved scales, 
involving physical processes (i.e., magnetohydrodynamics and microscopic transport processes) not
implemented in our runs. A fully self--consistent study of the
accretion onto the two MBHs down to few gravitational radii is beyond
the scope of this investigation.\\

 \indent
 The reason for the extreme variability in the earlier merger stage
(out of the two 
 described above) is
 twofold.  Firstly, the mass
accretion fluctuations are driven by the fast fluctuations in the
relative velocity between $M_2$ and the gas environment, that change
the $R_{\rm BHL}$ of $M_2$, and by the inhomogeneities in the gas that
passes
 in the vicinity of the secondary MBH.

Secondly, in the counter--rotating
 cases, the finite number of gas
particles that we use to sample the
 gas distribution may
occasionally lead to short quiescence periods
 when the secondary is
not accreting.  In such situations, the $R_{\rm
 BHL}$ of the
secondary is not resolved for a brief time due to the
 extremly high
relative velocity of the MBH with respect to the
 environment.  The
finite sampling implies also that values of
 $f_{\rm Edd}$ of order
of $10^{-2}$ should be considered order of
 magnitude estimates, and
can be incorrect by a factor of a
 few. However, values of $f_{\rm
Edd}$ larger than 0.1 have only a
 small relative numerical error of
$\lsim 10\,\%$.
Therefore, the general dependencies of the
luminosity variability on the initial merger parameters are robust.
A MBH orbiting in a circumnuclear disc during the final stages of
a galaxy merger is expected to be embedded in a gaseous/stellar
envelope comoving with the MBH itself. An accurate estimate of the
amount of the gas and stars embedding the retrograde MBH depends on
the initial conditions of the galaxy merger, and on the interplay
between different physical processes, e.g., tidal and ram--pressure
stripping. The presence of the stars and gas comoving with the MBH
would increase the mass that perturbs the circumnuclear disc and, as a
consequence, increase the effect of dynamical friction.  This may
reduce the timescale of the orbital angular momentum flip.  An envelope
of gas comoving with $M_2$ can also increase the accretion rate onto
the MBH, and decrease the variability of the accretion process, until
the envelope is not totally accreted by $M_2$ or removed by
gravitational/ram--pressure stripping processes.  We plan to study
such effects in a future paper.\\

\indent
The significant differences between $f_{\rm Edd}$ for MBH on eccentric
and circular orbits, and even stronger differences between $f_{\rm Edd}$ in
prograde and retrograde cases, can be considered a possible
explanation for the paucity of resolved double AGNs in circumnuclear
discs. More specifically, it is conceivable that a circumnuclear disk 
in a galaxy that underwent a merger may host a 
pair of MBH, but only one of them may 
show clear signatures of an AGN.
Alternatively, the secondary may be significantly dimmer, but highly variable.
Such scenarios may occur when the $M_2$ enters the disc on
a counter--rotating trajectory. Such effects have important implications for 
predicting the rates of double AGN detections in galaxies.\\
\indent
This study opens the way to predicting the properties of the accretion flows
near the horizons of the coalescing MBH. 
Our simulations predict the amount of gas delivered to the very vicinity of 
the inspiraling MBH. As such our simulations serve as 
initial conditions required to model the properties of the 
electromagnetic counterparts to the gravitational wave emission events 
that will be detectable with the {\it Laser
Interferometer Space Antenna (LISA)} (Armitage \& Natarajan 2002;
Kocsis et al. 2005; Milosavljevic \& Phinney 2005, Dotti et al. 2006a;
Lippai, Frei \& Haiman 2008; Schnittman \& Krolik 2008; Haiman, Kocsis
\& Menou 2008). We plan to improve upon our model by including gas
cooling and heating, star formation and supernova feedback from newly
formed stars. These new features can change the structure of the
circumnuclear disc, creating a less homogeneous multiphase environment
(see, e.g., Wada \& Norman 2001, Wada 2004). The interaction between
the two MBH and such a multiphase medium can in principle lead to a
larger eccentricity of the forming binary, and to more variable
accretion events onto the two MBH.

\section*{Acknowledgments}

 The authors acknowledge the anonymous referee for helpful comments.
MD wish to thank Roberto Decarli, David Merritt, Mark Reynolds,
Alberto Sesana, and Sandor Van Wassenhove for fruitful suggestions. MR
acknowledges {\it Chandra} Theory grant TM8-9011X. Support for this
work was provided by NASA grant NNX07AH22G and SAO-G07-8138 C (MV).

\end{document}